\documentclass[acmsmall]{acmart}

\setcopyright{cc}
\setcctype{by-nc-nd}
\acmJournal{PACMHCI}
\acmYear{2027}
\acmVolume{11}
\acmNumber{1}
\acmArticle{GROUP56}
\acmMonth{1}
\acmDOI{10.1145/3859442}

\copyrightyear{2027}

\received{31 March 2026}
\received[revised]{17 July 2026}
\received[accepted]{17 September 2026}

\AtBeginDocument{%
  }

\graphicspath{{figures/}}
\usepackage{url}
\usepackage{multirow}

\title[Who Shows Up, Who Benefits, and Who Wins?]{Who Shows Up, Who Benefits, and Who Wins? Expert-Novice Disparities in Team-Seeking Forums and Competition Performance on Kaggle}

\author{Ran Bi}
\orcid{0000-0001-5183-4601}
\affiliation{%
  \institution{SAS Institute Inc.}
  \city{Cary}
  \state{NC}
  \country{USA}}
\email{ran.bi@sas.com}

\author{Shiyao Wei}
\orcid{0000-0002-0844-0372}
\authornote{Corresponding author: \nolinkurl{sw22b@fsu.edu}.}
\affiliation{%
  \institution{Florida State University}
  \city{Tallahassee}
  \state{Florida}
  \country{USA}}
\email{sw22b@fsu.edu}

\renewcommand{\shortauthors}{Ran Bi and Shiyao Wei}

\begin{abstract}
Team formation is the first social gate to collaboration in online communities, yet it remains unclear whether public team-formation mechanisms broaden access to collaborators or reproduce existing hierarchies.
We examine this question on Kaggle, a data science competition platform whose ``Looking for a Team'' (L4T) megathreads let participants publicly seek teammates for on-platform competitions.
Grounded in Communities of Practice theory, we conduct two complementary studies of 18 competitions held in 2023.
Study~1 analyzes the reply network among 671 megathread participants and finds that 89.3\% are low-tier users, yet higher-tier users receive significantly more replies (Kruskal-Wallis $H=16.46$, $p=.00246$) and achieve higher rates of forum-visible teaming success (52.9\% vs.\ 32.6\%; $\chi^2=10.53$, $p=.001$), a proxy for eventual team membership.
Linking megathread posters to verified team rosters from 16 competitions partially supports this proxy. Twenty-six final teams contained two or more megathread posters, indicating that megathread interaction sometimes, but not always, maps onto realized teaming.
Study~2 links participation traces from 219 users to private leaderboard outcomes and finds that productive participation (submission count $\rho=.435$, cross-competition participation $\rho=.247$) is positively associated with performance, whereas megathread-specific discussion is not ($\rho=-.021$).
The submission-performance association appears stronger in higher tiers, although the novice subgroup estimate is imprecise.
Together, the findings suggest that open access to team-seeking forums does not guarantee equal participation. Visible expertise shapes who attracts attention, and engagement in the community's core practice is more strongly linked to success than discussion alone.
We discuss implications for designing more equitable team formation in online collaboration environments.
\end{abstract}



\begin{document}
\maketitle

\section{Introduction}
In many online communities, collaboration begins with a search for teammates. Members need to find teammates among strangers, relying on whatever public information the platform makes visible~\cite{harris:groupformation, marlow:impression}.
We study this process on Kaggle, a data science competition platform where individuals and teams compete by iteratively building predictive models and submitting solutions, often dozens of times per competition, to be ranked on competition leaderboards~\cite{twyman:kaggle, li:kaggle}.
Medals are awarded for strong competition placements and for well-received notebooks and datasets, and the visible performance tiers (Novice, Contributor, Expert, Master, Grandmaster), the top three of which require accumulated medals, are displayed throughout the platform.

Beyond competition, Kaggle has evolved into an informal learning space for professional development and collaboration~\cite{chow:kaggle, polak:kaggle}. Participants form teams, share notebooks, and discuss strategies, activities that mirror the social dynamics of Communities of Practice (CoP)~\cite{wenger:cop,cheng2020building}. In a CoP, members learn by taking part in a shared practice, and newcomers move from the periphery toward fuller participation~\cite{lave:lpp}.
To support team formation, Kaggle introduced ``Looking for a Team'' (L4T) megathreads, which are competition-specific forums where participants seek teammates~\cite{wei:megathread}.
These threads transform teammate search into a public process.
Instead of relying only on prior acquaintances, participants can advertise themselves and evaluate others.
That visibility creates opportunity, but it also turns reputation, prior achievements, and self-presentation into immediate sorting mechanisms~\cite{huang:crowdsourcing, harris:groupformation}.

These mechanisms raise two questions addressing equity and effectiveness.
First, who benefits from platform-provided team formation? Research on online learning at scale shows that designs and interventions that appear universally accessible do not benefit all learners equally~\cite{kizilcec:equity}.
We investigate whether analogous dynamics operate in open team-formation forums, where visible expertise may stratify outcomes.

Second, does community participation translate into measurable competitive success?
Research on online participation often struggles to connect social interaction to consequential outcomes~\cite{siemens:la} because self-reports are subjective, while forum activity may capture engagement without revealing whether participants gain practical advantage.
Kaggle's leaderboard offers an objective proxy, where rankings reflect demonstrated problem-solving ability on available data.

Cues such as tier badges, prior performance, and public activity histories can lower coordination costs during teammate search, but they can also stratify opportunity by making some forms of expertise more legible than others~\cite{harris:groupformation, marlow:impression}.
Team search serves different purposes for different users.
For established competitors, team search may be a way to form stronger teams or broaden technical coverage.
For students, career changers, and newer platform participants, it may also be a gateway into collaboration itself.
If the megathread systematically advantages already-visible users, then a seemingly open design may reproduce hierarchy at precisely the point where access matters most.

Prior work has described the megathreads through reply-network visualization and topic modeling~\cite{wei:megathread} and reported a tier-stratified analysis of the reply network and of final team membership~\cite{bi2026benefits}; other work has analyzed Kaggle's broader collaboration networks~\cite{twyman:kaggle, li:kaggle}.
However, no study has jointly examined who uses team-seeking forums, how interactions unfold across expertise levels, and whether taking part in team formation predicts competitive success.
This gap matters because team-formation mechanisms and competition outcomes are likely interdependent. Users who form more effective teams may also perform better in competitions.
We address this gap with two complementary studies using the same 2023 Kaggle dataset.

\begin{itemize}
\item \textbf{Study~1} examines the social network structure of megathread interactions, testing for disproportionate attention and expertise-based disparities in forum-visible teaming success.
\item \textbf{Study~2} links megathread participation to private leaderboard rankings, testing which participation dimensions predict performance and whether expertise moderates this relationship.
\end{itemize}

Kaggle is a well-suited natural laboratory for these questions~\cite{chow:kaggle, polak:kaggle}.
Unlike MOOCs, where participation may be driven by external requirements, Kaggle users are intrinsically motivated to perform well.
Private leaderboard rankings provide objective, standardized performance measures rather than subjective or instructor-assessed outcomes.
Finally, unlike laboratory team-formation studies, megathreads involve real-world collaborations with authentic motivations.
The same properties make Kaggle analytically demanding. Visible status cues may shape who gets noticed, performance metrics can be mistaken for learning writ large, and observed interaction does not guarantee that collaboration actually occurred.
Our analytic strategy therefore combines behavioral trace data with conservative interpretation, examining observable interaction patterns and linking them to participants' performance.

Together, these studies address four research questions:
\begin{itemize}
\item \textbf{RQ1:} How does expertise relate to megathread participation and network position?
\item \textbf{RQ2:} Do expert and novice users achieve forum-visible teaming success equally through megathread participation?
\item \textbf{RQ3:} Which participation dimensions predict competition performance?
\item \textbf{RQ4:} Does expertise moderate how participation relates to performance?
\end{itemize}

This paper makes three contributions.
Empirically, it provides a joint analysis of a public team-seeking channel that connects who participates, how attention is distributed across expertise levels, and whether participation is associated with competitive outcomes, by combining reply networks, verified team rosters, and private leaderboard rankings.
Theoretically, it speaks back to Communities of Practice theory by showing that, in platform-mediated communities with visible reputation systems, access to the periphery is graded rather than shared.
This qualifies the assumption that open access by itself sets newcomers on a path toward fuller participation.
For design, it separates social coordination from productive practice, showing that the two have different relationships with performance, and derives implications for building more equitable team-formation mechanisms.

\section{Theoretical Framework}
We ground our analysis in the framework of Communities of Practice (CoP)~\cite{lave:lpp}. The theory conceptualizes learning as a process of social participation within a community. In the framework, novices enter a community through legitimate peripheral participation~\cite{lave:lpp}, starting with observation and basic engagement before moving toward full participation and the center of the community. Kaggle competitions operate as self-directed learning environments where participants share knowledge, collaborate, and practice~\cite{marsick:informal}. We therefore treat the competition and the Kaggle community, more broadly, as the community of practice, and ask how novices gain access to it and whether that access translates into improving performance. 

Within this framework, we focus on two key mechanisms, namely access to community collaboration and the relationship between participation and performance. One of the central claims of legitimate peripheral participation is that access enables participants to move towards the center. The megathread provides such access, giving novices an open, public place to find teammates and begin collaborating. Another main argument of CoP is that participation is not just a social act but also a learning process. In the context of Kaggle, this means that the type and quality of participation (e.g., discussion vs.\ submission) may have different implications for learning and performance outcomes.

Lave and Wenger~\cite{lave:lpp} emphasize the importance of access to the learning community, suggesting that participation in forums like L4T can be a critical step for newcomers to gain access to the community and its resources. However, the theory also acknowledges that social structures and power dynamics within a community can influence who benefits from participation~\cite{lave:lpp}. In the context of Kaggle, visible expertise signals such as performance tiers may create barriers for novices, as they may feel intimidated by more experienced users or perceive that they have less to offer in terms of credibility~\cite{ren2020show}. This dynamic can lead to disparities in who benefits from the platform's collaborative mechanism, as established users may have an advantage in attracting attention and forming teams. Study 1 examines the team formation interaction in the megathread through network position, reply attention, self-presentation, and forum-visible teaming success.

In addition, beyond access, ongoing practices in the community contribute to learning and better performance~\cite{lave:lpp}. Once users are active in a competition, some forms of participation translate into competitive performance more than others. Study 2 examines this stage by relating discussion behavior and productive activity to leaderboard outcomes.
Section~\ref{sec:related} situates these two mechanisms in prior empirical work on Kaggle as a community of practice, on team formation and network dynamics (access), and on self-presentation and the participation--performance link.

\section{Related Work}
\label{sec:related}
\subsection{Kaggle as a Community of Practice}
Lave and Wenger's Communities of Practice framework describes learning through social participation, with newcomers entering via legitimate peripheral participation~\cite{lave:lpp}. In crowdsourcing contests, collaboration enhances the quality of submissions and ideas~\cite{blohm2011does}, and participants are motivated to contribute to both their teams and the broader community~\cite{javadi2024contribution}. Kaggle extends these ideas into a competitive, self-directed environment in which learning is embedded in repeated model building, feedback interpretation, and public comparison. This places Kaggle closer to informal workplace collaboration than to instructor-led discussion spaces~\cite{marsick:informal}. While research on online courses finds that how closely learners read peers' forum posts relates to the quality of their own posts~\cite{wise:participation}, such findings may not transfer cleanly to voluntary competition settings where objective performance signals and public reputation cues are central. Furthermore, competitive platform designs impact experts and novices differently. Prior work indicates that while experts share knowledge to compete for reputation, beginners are often driven away from public contribution due to their vulnerable social image~\cite{cheng2020building}. Our study extends this observation to team formation, examining how the competitive environment of Kaggle's megathreads affects newcomers attempting to establish collaborative relationships without established platform credentials.

Because our questions concern how this environment shapes team formation, the platform's design details matter. Each Kaggle competition provides a dataset, an objective evaluation metric, and a deadline; participants may submit predictions repeatedly, receiving provisional feedback on a public leaderboard while final rankings are computed on a held-out private leaderboard. Strong placements earn medals, as do well-received notebooks and datasets; accumulated medals confer the higher performance tiers, and a badge for the user's overall tier, the highest reached in any track, is displayed beside the user's name in every forum post and profile. Users may compete solo or merge into teams before a competition-specific deadline, and each competition hosts a discussion forum and shared notebooks alongside the leaderboard. The L4T megathread sits inside this infrastructure as an open, unstructured matching channel. Anyone can post, no algorithm recommends candidates, and every request is read against the poster's visible reputation.

\subsection{Team Formation and Network Dynamics}
Group formation is the first organizing step in online collaboration~\cite{harris:groupformation}. 
In self-formed teams, personal agency is high because participants decide whom to contact, what information to reveal, and whether potential collaborators appear trustworthy or useful.
In Kaggle, this process is driven by public activity traces, visible expertise markers, and communication cues~\cite{marlow:impression}. Research on crowdsourcing competitions has examined how homophily and diversity in solvers' demographic and acquired characteristics relate to teammate choice and team performance~\cite{dissanayake:crowdsourcing}, and past collaboration histories can also shape team-up decisions in crowdsourcing contests~\cite{huang:crowdsourcing}.
Because network positioning matters for performance in Kaggle competitions~\cite{li:kaggle,twyman:kaggle}, network mechanisms like preferential attachment~\cite{barabasi:preferential} and homophily~\cite{mcpherson:homophily} are not only structural concepts but also equity-relevant mechanisms. If experienced users attract disproportionate attention, open team-seeking channels may still reproduce unequal access.

This reliance on self-formed teams also differs from the team-formation mechanisms most often studied in social computing, which typically add structure or mediation to matching. For example, Huddler forms crowd teams from workers who have collaborated before, even when their availability is unpredictable~\cite{salehi2017huddler}. Team dating lets strangers try short collaborations before they commit to a team~\cite{lykourentzou2017team}. Recommender systems help users search for and select teammates based on profile attributes~\cite{gomez2019would}, and digital nudges steer self-formed crowd teams toward better compositions~\cite{vinella:nudging}.
Kaggle's megathread offers none of that scaffolding. Matching is unassisted, and public reputation cues are present at every step.
Findings from the megathread should therefore transfer most directly to platforms where members choose collaborators themselves under visible reputation systems, and less directly to platforms that mediate matching algorithmically.

\subsection{Self-presentation and Trust Signals}
Group formation depends on the information available about prospective members. A review of CSCW research on group formation from 1992 to 2018 identifies self-presentation as one of six enduring themes, with information availability as one of its sub-themes~\cite{harris:groupformation}. When collaboration begins among strangers, participants engage in impression formation, making rapid judgments about competence and reliability from whatever cues are at hand, such as activity traces, personal profiles, and selectively disclosed information~\cite{marlow:impression}. In online peer production, where direct acquaintance with collaborators is rare, these public traces do much of the work that prior familiarity does offline~\cite{marlow:impression}. Crucially, not all cues carry equal weight. Warranting theory holds that observers discount claims a presenter can cheaply make about themselves and place more trust in cues that are difficult to manipulate, such as system-generated records of past behavior~\cite{walther:warranting}. Consistent with this, Ren et al.\ find that open-innovation participants who share online profiles perform better and are more likely to become team leaders, and that this advantage runs through team formation. Profile sharers are better able to recruit high-performing teammates, and that formation effectiveness in turn accounts for the link between self-disclosure and success~\cite{ren2020show}.

On Kaggle, this profile-viewing step is built into the megathread interface. A reader evaluating a team-seeking post sees the poster's tier badge beside the username and is one click away from a platform-authored profile of medals, tenure, and competition history, so every request for teammates effectively arrives with a profile attached. Unlike the profiles in Ren et al.'s setting, however, the most salient element of this profile is compiled by the platform rather than curated by the user. The tier system compresses a long competitive history into an immediately readable badge, a system-generated warrant that helps participants make fast decisions in a crowded environment~\cite{ren2020show, walther:warranting}. This efficiency is useful for teammate search, but it may also introduce bias into the formation process. If voluntarily shared profiles improve team formation for users whose records are strong~\cite{ren2020show}, a profile that accompanies every post by default may work against users whose records are thin, narrowing the set of users deemed worth replying to. Research shows that while users initially search for experts, they often default to prior social connections, leading to segregated teams where expertise and social capital are highly concentrated~\cite{gomez2019would}. Evidence from entrepreneurial founding teams likewise suggests that novices and experienced founders weigh different criteria when choosing team members~\cite{das2023comparative}. For newcomers, system-generated warrants are sparse. Lacking medals and badges, they must establish credibility through self-generated claims such as motivation statements, contact sharing, or assertions of transferable experience, precisely the cues warranting theory predicts will be discounted. By analyzing Kaggle's L4T megathreads, we investigate how the platform's visible tier system functions as an information signal, shaping who receives attention and how trust is negotiated.

\subsection{Participation and Performance}
Beyond initial team formation, researchers have investigated how team structure and individual positions impact long-term performance. In crowdsourcing contests, the choice between solo work and transient teamwork is itself consequential and path-dependent. Prior teamwork builds team-related capital that makes future collaboration more likely, while accumulated solo experience discourages it, so early collaboration experiences compound over a contestant's history~\cite{huang:crowdsourcing}. This path dependence makes the team-formation stage we study especially important, because disparities in who can form a first team may propagate into long-run differences in how people participate. A user's positioning within a collaboration network, such as their number of collaborations and diversity of experience, has a positive relationship with the medals and points they earn~\cite{twyman:kaggle}. While these studies establish that team formation and network positioning matter, they leave open whether the behaviors involved in finding a team differ from those involved in doing the competition work. On Kaggle, the core practice is around building and submitting models but not around finding a teammate. Each submission is a specific engagement with the problem. Posting in a team-seeking thread, by contrast, is peripheral social coordination rather than participation in the practice. From a Communities of Practice standpoint, this is the difference between being socially present at the periphery and engaging in the activity through which competence develops.

Accordingly, there is a need to distinguish between the social act of finding a team and the productive act of competing. We build on prior literature investigating ad hoc teams~\cite{salehi2017huddler, lykourentzou2017team} by distinguishing discussion-based participation from productive participation, and linking these specific dimensions directly to private leaderboard outcomes. This distinction is methodologically important because studies of online participation often aggregate posting, viewing, and producing into broad engagement measures, masking whether behaviors reflect coordination or actual skill practice. Prior work on the L4T megathreads themselves consists of two posters. One visualizes reply networks and uses topic modeling to characterize what team-seeking posts say~\cite{wei:megathread}; the other examines expert--novice differences in online team formation~\cite{bi2026benefits}. We extend this work by linking forum traces to private leaderboard outcomes (Study~2) and interpreting both studies through Communities of Practice, so that we can test which behaviors are associated with competitive success rather than only describe them.

\section{Data and Methods}
\subsection{Data Collection}

We collected data from Kaggle's L4T megathreads for competitions held in 2023 using the Kaggle API.
The linkage strategy followed the natural structure of the platform.
We first collected competition-level megathread comments and reconstructed reply relations from comment metadata.
We then merged those records with public user-profile information, private leaderboard outcomes where available, and final team rosters for competitions with compatible outcome data.
This design preserves the social context of megathread participation while allowing us to connect forum traces to downstream competition outcomes.
Study~2 is limited to the 16 competitions with US-dollar prizes; the other two were retained in Study~1 because their interaction data were still informative for network analysis.
Our dataset comprises four linked components.

\textbf{Megathread comments.} 1,577 comments from 676 unique users across 18 competitions.
Each record includes message text, reply structure, medal status, and competition identifier.
Competitions span NLP, computer vision, and tabular prediction.
We anonymized all user-level handling during analysis and report only aggregate results.

\textbf{User profiles.} 671 users matched to Kaggle's performance tier system (0=Novice, 1=Contributor, 2=Expert, 3=Master, 4=Grandmaster; 99.3\% match rate).
Kaggle also assigns a Staff designation (PerformanceTier=5) to platform employees and moderators; a small number of Staff accounts appear in the dataset but are excluded from all tier-stratified analyses because their engagement pattern (e.g., pinned announcements that attract many replies) differs structurally from competitive participation.
We create a binary split for analysis into Low-Tier (tiers 0--1; 89.3\%) and High-Tier (tiers 2--4; 10.4\%) users; the two Staff accounts (0.3\%) are excluded.
Kaggle's tier system is based on accumulated medals. The Expert, Master, and Grandmaster tiers require medals, which are awarded for competition placements and for well-received notebooks and datasets, and the overall tier shown on a profile is the highest tier reached in any track. This makes tier a reasonable proxy for demonstrated data science expertise. We use each user's overall tier as recorded in Meta Kaggle, Kaggle's public metadata dataset, in early January 2024, not the tier displayed when a comment was posted.
However, tier is a cumulative, cross-competition achievement signal rather than a direct measure of performance in a focal competition.
The system also conflates skill with platform tenure, as newer users may have substantial real-world expertise not yet reflected in their Kaggle badge.
Accordingly, we interpret tier as a practically salient but imperfect marker. It is visible to other users during teammate search, but it does not by itself determine how well someone will perform in a new competition.

\textbf{Competition performance.} For 219 users who both posted in megathreads and submitted to competitions, we obtained private leaderboard rankings across the 16 competitions with US-dollar prizes, yielding 382 user-competition observations.
These observations form the basis of the participation-performance analysis and allow the same user to appear in multiple competitions.

\textbf{Verified team rosters.} To assess whether megathread interactions correspond to realized teams, we linked 219 megathread posters to final team rosters for the same 16 reward competitions used in Study~2.
This validation file contains 1,199 post-to-team records spanning 342 teams.
Of these teams, 105 had more than one member, and 26 included two or more megathread posters.
These verified overlaps do not imply that every reciprocal reply became a team, but they establish that megathread participation sometimes translated into observable same-team membership.
The unit of this validation file is the post-to-team record rather than the unique user or unique team.
Accordingly, we use the file descriptively to assess how often megathread posters appear in realized teams and how often multiple posters from the same thread converge on a single team.

Table~\ref{tab:desc} summarizes the datasets.
Users in our sample registered between 2013 and 2023, with a median tenure of 1.9 years (mean 2.5) as of 31 December 2023.
The tier distribution is heavily skewed. Novices and Contributors together account for 89.3\% of the megathread population.

\begin{table}[t]
\centering
\small
\caption{Descriptive statistics for both studies. \emph{Teams} = total / multiplayer / with $\geq$2 L4T posters; \emph{Network} = nodes / edges / density.}
\begin{tabular}{@{}lr@{\hskip 2.5em}lr@{}} \toprule
\multicolumn{2}{c}{\textbf{Study 1 (Network)}} & \multicolumn{2}{c}{\textbf{Study 2 (Performance)}} \\
\cmidrule(r){1-2}\cmidrule(l){3-4}
Competitions     & 18                   & Competitions          & 16   \\
Unique users     & 671                  & Unique users          & 219  \\
Total comments   & 1{,}577              & User--comp.\ obs.\    & 382  \\
Post--team links & 1{,}199              & Mean submissions/user & 28.8 \\
Teams            & 342 / 105 / 26       & Mean team size        & 1.8  \\
Network          & 665 / 1{,}300 / .003 &                       &      \\
\bottomrule
\end{tabular}
\label{tab:desc}
\end{table}

\subsection{Study 1: Network Analysis of Megathread Interactions}

Study~1 addresses RQ1 and RQ2, both of which concern how expertise shapes interaction \emph{inside} the megathread.
A reply network is the natural instrument because it lets us measure who attracts attention and occupies central positions (RQ1) and whether that attention converts into forum-visible teaming success (RQ2), using observed interaction traces rather than self-reports.
We constructed a directed reply network using NetworkX~\cite{hagberg:networkx}, in which each node is a user and a directed edge from $A$ to $B$ indicates that $A$ replied to a comment by $B$.
The network contains 665 nodes and 1,300 edges (density = 0.003).
Because interaction and activity measures are highly skewed, we rely primarily on nonparametric comparisons and descriptive network summaries rather than distribution-sensitive mean comparisons alone.

\textbf{Network metrics (RQ1).} We computed in-degree (replies received), out-degree (replies sent), and betweenness centrality per node, comparing distributions across tiers using Kruskal-Wallis tests.

\textbf{Disproportionate attention (RQ1).} We compared the proportion of replies received by each tier against their population share, tested with chi-squared goodness-of-fit. This is a cross-sectional test of whether attention is distributed in proportion to each tier's representation, not a temporal test of cumulative-advantage growth.

\textbf{Cross-tier reply patterns (RQ1).} To describe how attention flows across expertise levels, we tabulated the reply mixing matrix, the share of each tier's replies directed to every other tier.
We treat this matrix as a descriptive summary of cross-tier interaction rather than as a formal homophily test.

\textbf{Forum-visible teaming success (RQ2).} Our primary Study~1 outcome is whether a user receives a visible signal that their megathread participation gained traction, operationalized as receiving a medal on a comment or having reciprocal reply interactions.
Because this measure does not itself verify eventual team membership, we treat it as a proxy rather than a direct roster outcome.
We partially validate it by linking megathread posters to verified team rosters for 16 competitions; the linkage identifies 342 teams represented by megathread posters, including 26 teams in which at least two members posted in the megathread.
This validation supports interpreting the metric as a meaningful but imperfect indicator of realized team formation.
Put differently, the proxy captures whether public interaction appears to move beyond one-sided solicitation, while the roster linkage shows how often those public traces can be connected to eventual same-team membership.

\textbf{Trust signal analysis (RQ2).} We identified five self-presentation categories via keywords (motivation, credentials, contact sharing, skills, and past performance).
The dictionaries were developed iteratively from close reading of the megathread corpus and the coding scheme used in prior exploratory megathread work~\cite{wei:megathread}.
We computed the proportion of messages in each tier containing keywords from each category.
Because the final classification is keyword-based rather than dual human coding, we treat these results as descriptive indicators of self-presentation rather than a fully validated content-analysis measure.
The five categories were chosen because they recur frequently in team-seeking discourse and correspond to distinct ways of claiming credibility.
Motivation captures willingness to learn or work hard; credentials capture occupational or educational identity; contact sharing captures invitations to continue off-platform; skills capture technical competencies such as Python, deep learning, or competition-specific domains; and past performance captures explicit references to prior placements or medals.
Table~\ref{tab:keywords} lists the keywords for all five categories.

\begin{table}[t]
\centering
\small
\caption{Keyword dictionary for the five self-presentation categories (Study~1). Matching uses lowercase substrings without word boundaries; any listed keyword triggers its category, and categories may overlap.}
\begin{tabular}{@{}lp{0.74\linewidth}@{}} \toprule
\textbf{Category} & \textbf{Keywords} \\ \midrule
Motivation       & learn, learning, beginner, new, first, start, improve, grow, practice, explore \\
Credentials      & degree, phd, master, bachelor, university, student, engineer, scientist, researcher, professor, experience \\
Contact sharing  & email, discord, linkedin, slack, gmail, telegram, whatsapp, skype \\
Skills           & python, machine learning, deep learning, tensorflow, pytorch, kaggle, notebook, model, neural, nlp, cv, data science \\
Past performance & gold, silver, bronze, medal, rank, top, scored, leaderboard, winner \\
\bottomrule
\end{tabular}
\label{tab:keywords}
\end{table}

\subsection{Study 2: Participation-Performance Analysis}

Study~2 addresses RQ3 and RQ4, which ask whether participation translates into \emph{competitive} performance rather than forum-visible success.
We therefore link participation traces to private leaderboard rankings, an outcome that is external to the forum and not defined in terms of tier, so that any association between participation and performance is not true by construction.
For each user-competition pair ($N=382$), we computed two sets of features.

\textbf{Participation metrics.} Discussion-based participation includes megathread comment count and message length, and productive participation includes submissions and cross-competition engagement.
We do not analyze replies made as a separate metric. Except for the 18 thread-opening posts, none of which were authored by Study~2 users, every megathread comment is a reply, so reply count is identical to comment count in this sample.
Discussion captures social engagement, submissions capture engagement in the community's core practice of building and evaluating models~\cite{lave:lpp}.

\textbf{Performance.} We use $100 - \text{percentile rank}$ on the private leaderboard (higher = better, range 0--100).
The leaderboard uses a held-out test set, revealed after competition close.
This measure is not equivalent to all forms of learning, but it provides a consistent, platform-native indicator of competitive problem-solving under comparable evaluation conditions.

\textbf{Analytical approach.} We compute Spearman correlations between participation metrics and performance (RQ3), then fit multiple linear regression with five-fold cross-validation~\cite{scikit-learn} as a robustness check on predictive stability.
Because the dataset contains repeated user-competition observations, these model estimates should be interpreted descriptively rather than as a fully user-independent generalization test.
For RQ4, we examine whether expertise changes the strength of the submission-performance relationship using tier-stratified Spearman correlations and a Fisher $z$-test comparing the extreme subgroups. Because the Novice and Master subsamples ($n=21$ each) are too small to support a well-powered interaction term, we treat these as descriptive tier-stratified associations rather than fitting a formal submission-by-tier interaction. We separately report a hierarchical regression that adds tier as a main effect; this tests whether visible expertise explains performance beyond participation---guarding against a circular ``experts win'' reading---rather than testing the moderation itself.
We use Spearman correlations because several participation measures are skewed and potentially nonlinear in relation to performance.
Taken together, the correlation and regression analyses let us ask both which traces matter in isolation and whether the same traces remain informative once other forms of participation are considered simultaneously.

\section{Results}

\subsection{RQ1: Expertise and Participation Patterns}

\begin{figure}[t]
\Description{Bar chart of megathread participant counts by performance tier: Novice 283 (42.2\%), Contributor 316 (47.1\%), Expert 62 (9.2\%), Master 7 (1.0\%), Grandmaster 1 (0.1\%), and Staff 2 (0.3\%). Low-tier users (Novice and Contributor) dominate.}
\centering
\includegraphics[width=0.6\columnwidth]{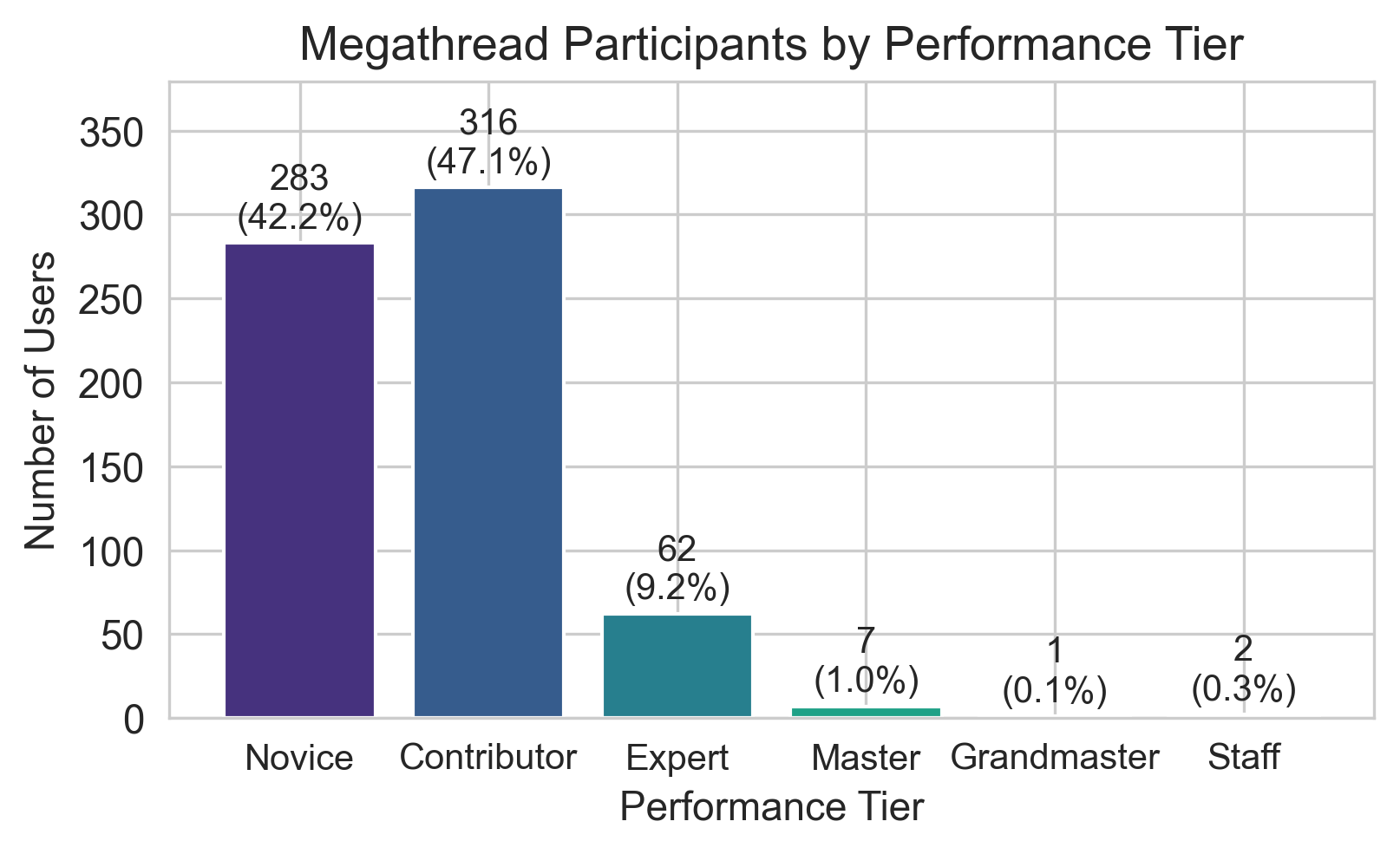}
\caption{Megathread participants by performance tier. Low-tier users (Novice and Contributor) dominate at 89.3\%, versus 10.4\% high-tier (Expert and above).}
\label{fig:tiers}
\end{figure}

We begin by establishing who participates in the megathread and how attention is distributed across the reply network.
The megathread population is heavily skewed toward lower tiers (Figure~\ref{fig:tiers}).
Novice users comprise 42.2\% and Contributors 47.1\%, while Expert users account for only 9.2\%, with Master (1.0\%) and Grandmaster (0.1\%) users nearly absent.

Despite their minority presence, higher-tier users occupy more prominent network positions (Table~\ref{tab:network}).
Reply attention differs significantly across tiers (Kruskal-Wallis $H=16.46$, $p=.00246$). Expert users receive the most replies (mean in-degree = 3.32), well above Novice (1.39) and Contributor (1.43) users.
Betweenness centrality also differs significantly ($H=16.09$, $p=.003$), with Experts serving as more central bridges in the reply network.
Out-degree does not differ ($H=3.82$, $p=.431$). All tiers send similar numbers of replies but receive attention unevenly, revealing an asymmetry in social capital accumulation.
This asymmetry is substantively important.
If users across tiers speak at roughly similar rates but only some tiers reliably attract responses, then inequality is emerging at the level of reception rather than effort.
Novice users are not absent from the conversation; instead, they participate in a forum where replies are distributed unevenly.

\begin{table}[t]
\centering
\small
\caption{Mean network metrics by tier (Study~1). In-degree and betweenness differ; out-degree does not.}
\begin{tabular*}{\linewidth}{@{\extracolsep{\fill}}lcccc@{}} \toprule
\textbf{Metric} & \textbf{Novice} & \textbf{Contributor} & \textbf{Expert} & \textbf{Master} \\ \midrule
In-degree   & 1.39  & 1.43  & 3.32  & 2.43 \\
Out-degree  & 2.20  & 2.17  & 2.87  & 2.00 \\
Betweenness & .0004 & .0003 & .0021 & .0014 \\
\bottomrule
\end{tabular*}
\par\smallskip
{\footnotesize\parbox{\linewidth}{Kruskal-Wallis $H=16.46$ for in-degree ($p=.00246$), 3.82 for out-degree ($p=.431$), and 16.09 for betweenness ($p=.003$). Grandmaster ($n{\leq}1$) and Staff omitted.}}
\label{tab:network}
\end{table}

\begin{figure}[t]
\Description{Heatmap showing reply mixing matrix between performance tiers. Left panel shows raw counts; right panel shows row-normalized percentages. The matrix reveals that novice and contributor users direct replies broadly, while expert users show more concentrated interaction patterns.}
\centering
\includegraphics[width=\textwidth]{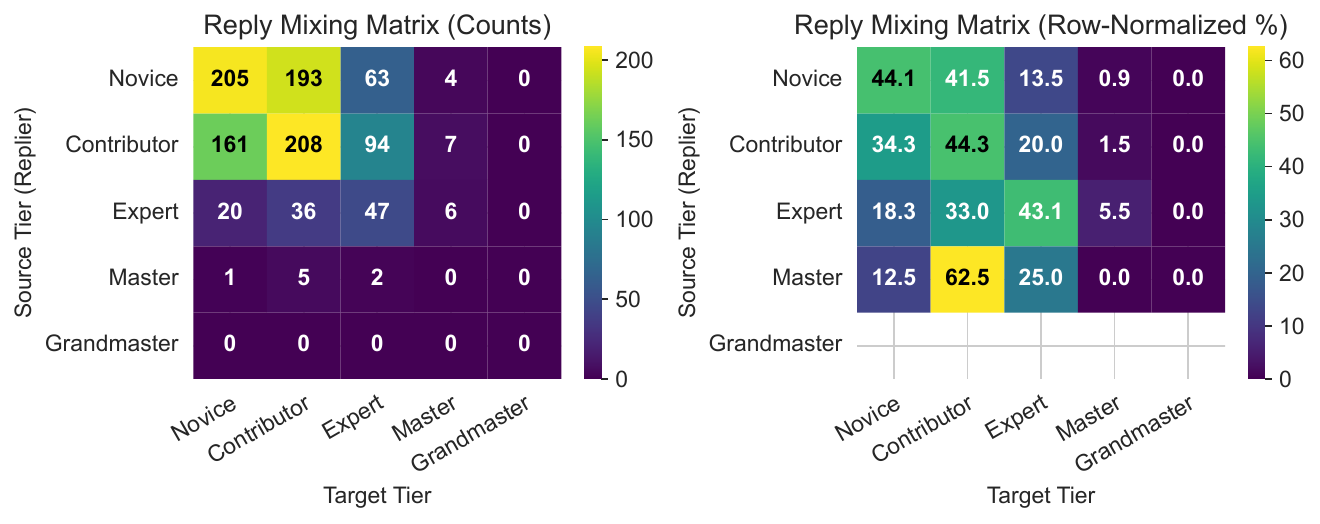}
\caption{Reply mixing matrix (Study~1). Novice and Contributor users direct many of their replies toward higher-tier targets, while Expert replies are more concentrated. The matrix is descriptive.}
\label{fig:homophily}
\end{figure}

\begin{figure}[t]
\Description{Network visualization showing the largest connected component of the megathread reply network. Nodes are colored by performance tier on a viridis scale (dark purple = Novice through yellow-green = Grandmaster), with Staff accounts shown in gray. Node size is proportional to in-degree. Higher-tier nodes appear more centrally positioned with more incoming connections.}
\centering
\includegraphics[width=0.95\textwidth]{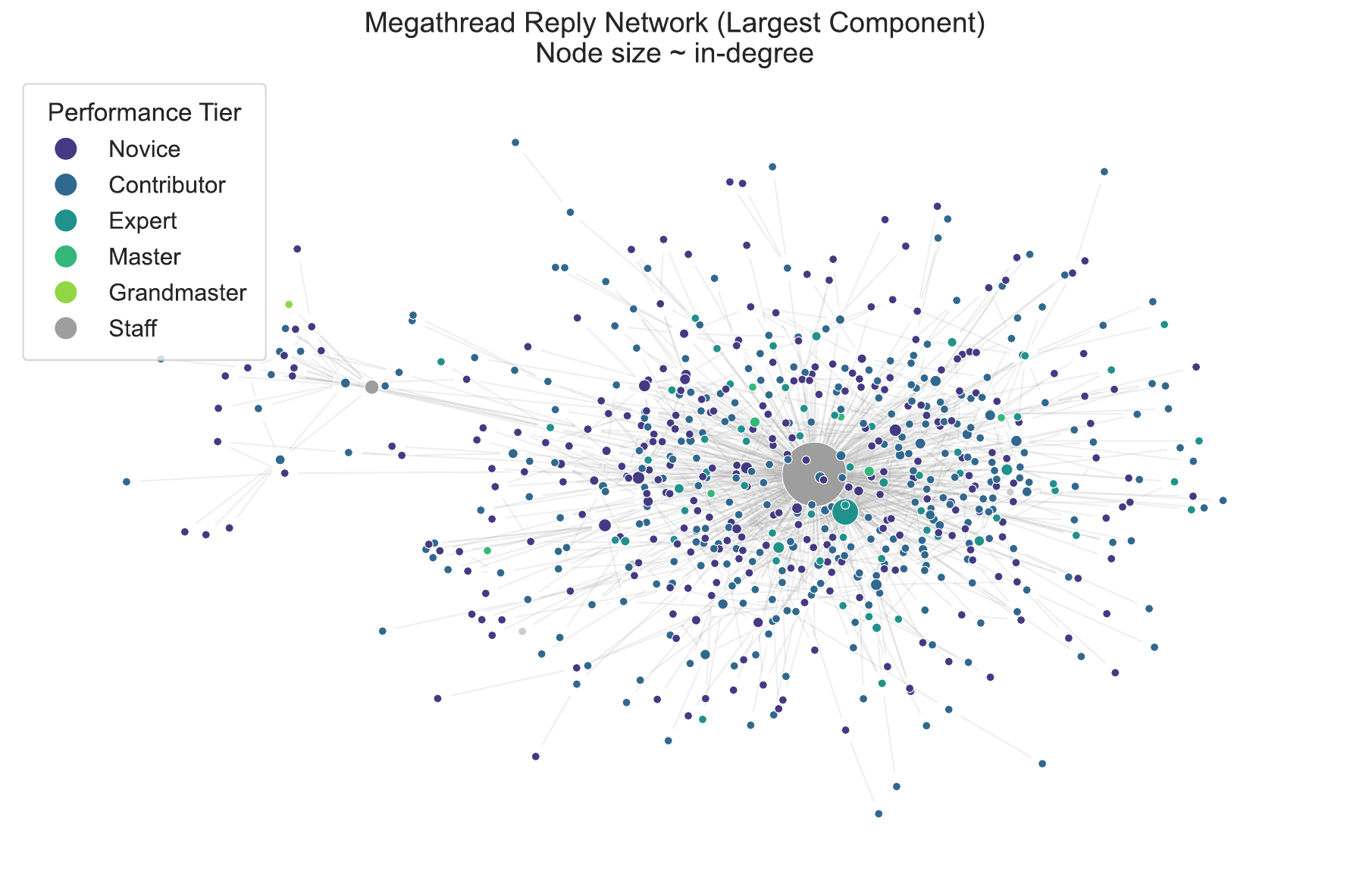}
\caption{Megathread reply network (665 nodes). Node size=in-degree; color=tier. Higher-tier users tend to occupy central positions.}
\label{fig:network}
\end{figure}

Experts receive $1.51\times$ their population share in replies, while Novices ($0.62\times$) and Contributors ($0.63\times$) receive less than expected (chi-squared $p<.001$; Table~\ref{tab:pref}), indicating that higher-tier users attract a disproportionate share of attention.
Consistent with this upward flow of attention, the reply mixing matrix (Figure~\ref{fig:homophily}) shows that Novice and Contributor users direct many of their replies toward higher-tier targets, whereas Expert replies are more concentrated.
Figure~\ref{fig:network} visualizes the reply network, in which higher-tier nodes tend to occupy more central positions.

\begin{table}[t]
\centering
\small
\caption{Population share vs.\ reply-target share by tier (Study~1, $N=665$ nodes). A ratio above 1 indicates disproportionate attention.}
\begin{tabular*}{\linewidth}{@{\extracolsep{\fill}}lcccc@{}} \toprule
\textbf{Metric} & \textbf{Novice} & \textbf{Contributor} & \textbf{Expert} & \textbf{Master} \\ \midrule
Population \%    & 42.2 & 47.1 &  9.2 &  1.0 \\
Reply-target \% & 26.2 & 29.9 & 13.9 &  1.1 \\
Ratio           & 0.62 & 0.63 & 1.51 & 1.10 \\
\bottomrule
\end{tabular*}
\par\smallskip
{\footnotesize\parbox{\linewidth}{Competitive tiers only. Staff (Kaggle moderators) and Grandmaster accounts are omitted; the remaining ${\sim}28.9\%$ of reply-targets is attributable primarily to Staff accounts, whose pinned megathread announcements attract many replies.}}
\label{tab:pref}
\end{table}

\subsection{RQ2: Differential Forum-Visible Success Rates}

\begin{figure}[t]
\Description{Bar chart showing success rate by tier: Novice 32.5 percent, Contributor 32.6 percent, Expert 51.6 percent, Master 57.1 percent. Higher tiers show progressively better success rates.}
\centering
\includegraphics[width=0.6\columnwidth]{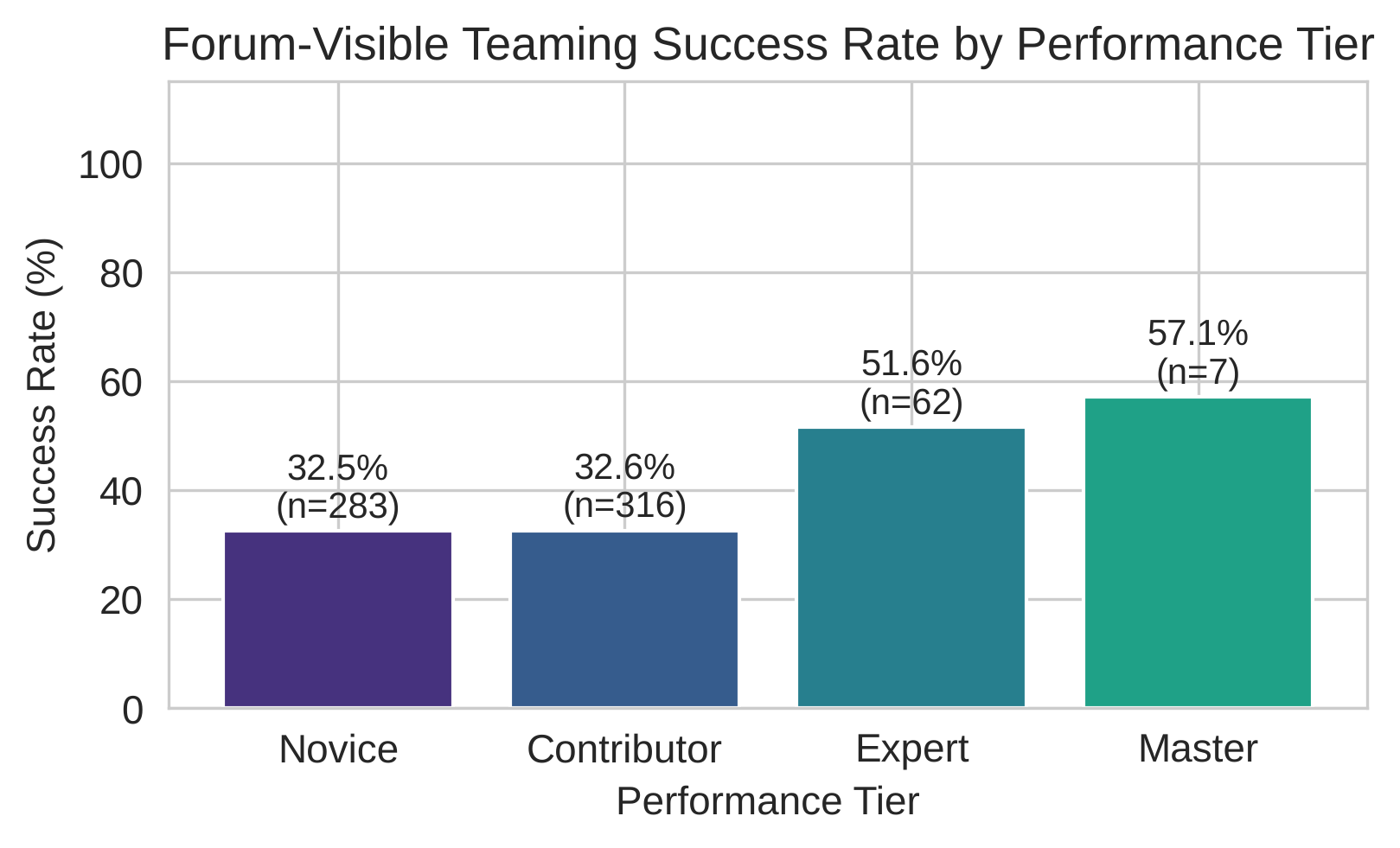}
\caption{Forum-visible teaming success rate by tier (Study~1). High-tier users achieve 52.9\% vs.\ 32.6\% for low-tier users ($p=.001$).}
\label{fig:success}
\end{figure}

RQ2 stays within the megathread and asks whether the attention documented under RQ1 translates into forum-visible teaming success, a Study~1 outcome that is internal to the forum and distinct from the leaderboard performance examined in Study~2 (RQ3--RQ4).
Forum-visible teaming success differs significantly by expertise.
High-tier users (Expert+) achieve a 52.9\% success rate compared to 32.6\% for low-tier users ($\chi^2=10.53$, $p=.001$; Figure~\ref{fig:success}).
Trust signal analysis (Table~\ref{tab:trust}) reveals systematic differences. Novices emphasize motivation (30.3\%) and contact sharing (21.3\%), whereas explicit references to past performance are rare in every tier, rising from 1.2\% of Novice messages to 13.3\% of Master messages (2 of 15).
Novices appear to rely more heavily on low-cost credibility signals, whereas higher-tier users can point to prior performance that is already legible on the platform.
Roster validation shows that the proxy captures some, but not all, realized teaming.
Across the 16 linked competitions, 219 megathread posters were associated with 342 final teams; 105 of those teams were multiplayer teams, and 26 verified teams included at least two megathread posters.
These overlaps indicate that megathread participation sometimes culminated in actual same-team membership, but they also bound how far the proxy can be trusted, because only 26 of 105 multiplayer teams (and a small fraction of all 342 linked teams) contained two or more megathread posters.
We therefore read the roster linkage as suggestive, partial validation rather than confirmation that forum-visible success equals realized teaming.

\begin{table}[t]
\centering
\small
\caption{Trust signal usage rates (\%) by tier. Novices emphasize motivation and contact sharing; references to past performance are rare in all tiers.}
\begin{tabular*}{\linewidth}{@{\extracolsep{\fill}}lcccc@{}} \toprule
\textbf{Signal Category} & \textbf{Novice} & \textbf{Contrib.} & \textbf{Expert} & \textbf{Master} \\ \midrule
Motivation       & 30.3 & 28.3 & 19.0 & 20.0 \\
Credentials      & 25.5 & 30.7 & 14.1 & 20.0 \\
Contact sharing  & 21.3 & 19.1 & 10.3 &  6.7 \\
Skills           & 32.0 & 30.9 & 25.5 & 26.7 \\
Past performance &  1.2 &  1.8 &  3.8 & 13.3 \\
\bottomrule
\end{tabular*}
\par\smallskip
{\footnotesize\parbox{\linewidth}{Grandmaster ($n{=}1$) and Staff omitted.}}
\label{tab:trust}
\end{table}

\subsection{RQ3: Participation and Performance}

\begin{figure}[t]
\Description{Two scatter plots of competition performance against submission count (left) and megathread comment count (right). Submission count has a positive association with performance, while megathread comment count has little association. Both panels retain substantial variance.}
\centering
\includegraphics[width=\textwidth]{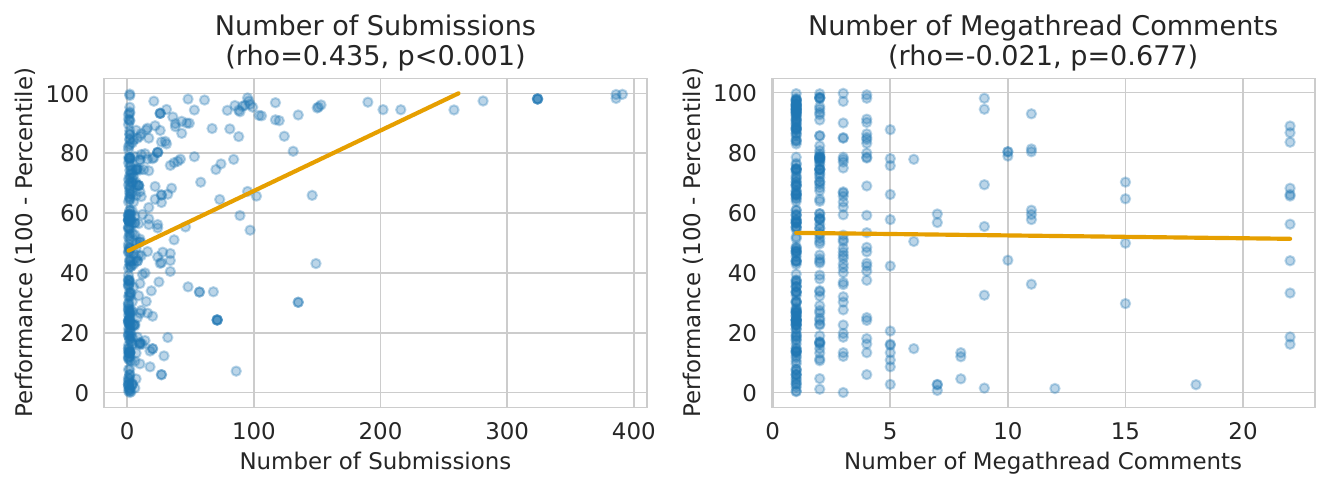}
\caption{Competition performance versus submission count (left) and megathread comment count (right; Study~2, $N=382$). More submissions are associated with better leaderboard placement ($\rho=.435$, $p<.001$), whereas megathread comments are not ($\rho=-.021$, $p=.677$).}
\label{fig:scatter}
\end{figure}

\begin{figure}[t]
\centering
\includegraphics[width=0.6\columnwidth]{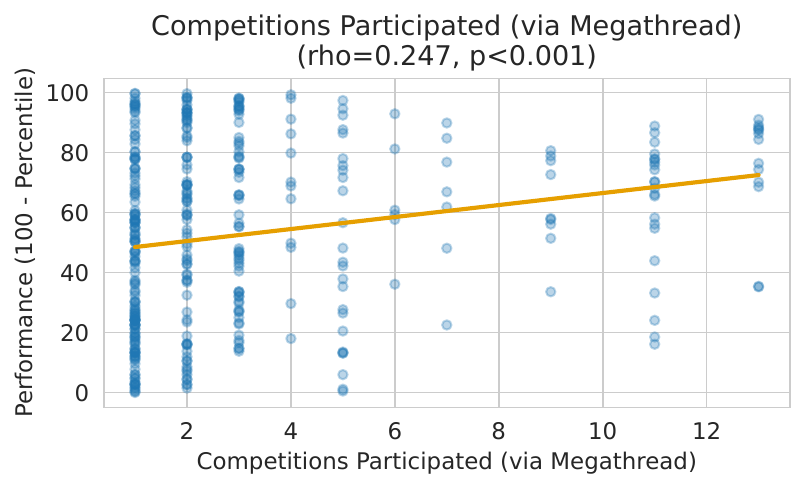}
\caption{Competition performance versus cross-competition engagement ($\rho=.247$; Study~2, $N=382$).}
\Description{Scatter plot of the Study 2 data showing that performance is positively associated with cross-competition engagement.}
\label{fig:scatter-context}
\end{figure}

We now move from the forum to the leaderboard, asking which dimensions of participation track competitive performance (Study~2).
Linking participation to leaderboard outcomes (Table~\ref{tab:corr}), submission count shows the strongest correlation with performance ($\rho=.435$, $p<.001$; Figure~\ref{fig:scatter}, left), followed by cross-competition engagement ($\rho=.247$; Figure~\ref{fig:scatter-context}).
In contrast, megathread-specific discussion shows no relationship with performance (comment count $\rho=-.021$, $p=.677$; Figure~\ref{fig:scatter}, right).

\begin{table}[t]
\centering
\small
\setlength{\tabcolsep}{4pt}
\caption{Spearman correlations between participation and competition performance (Study~2, $N=382$).}
\begin{tabular*}{\linewidth}{@{\extracolsep{\fill}}lcccc@{}} \toprule
 & \textbf{Submission} & \textbf{Cross-comp.} & \textbf{Avg.\ msg.} & \textbf{Megathread} \\
 & \textbf{count} & \textbf{engagement} & \textbf{length} & \textbf{comments} \\ \midrule
$\boldsymbol{\rho}$ & $+$.435 & $+$.247 & $-$.119 & $-$.021 \\
$\boldsymbol{p}$    & $<$.001 & $<$.001 & .020    & .677    \\
\bottomrule
\end{tabular*}
\label{tab:corr}
\end{table}

\begin{figure}[t]
\Description{Horizontal bar chart of regression coefficients on four standardized predictors. Submission count has the largest positive bar at 12.55, followed by cross-competition engagement at 7.90. Megathread comments (-2.41) and average message length (-1.29) show small negative bars. Values represent expected change in performance (0--100 scale) per one-SD increase in the predictor.}
\centering
\includegraphics[width=0.67\columnwidth]{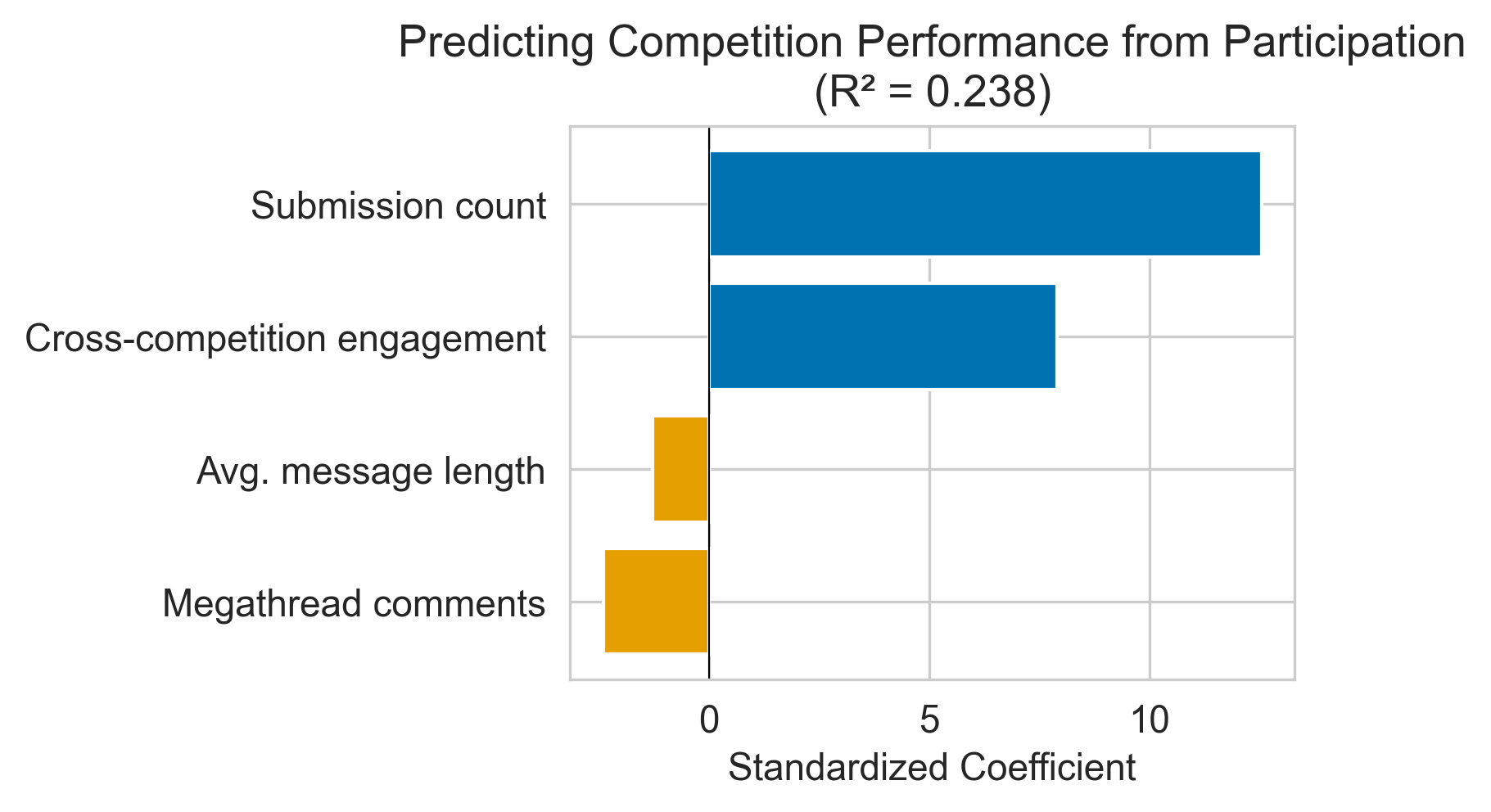}
\caption{Standardized regression coefficients predicting performance (Study~2, $R^2=.238$).}
\label{fig:coefs}
\end{figure}

A multiple regression explains 23.8\% of variance ($R^2=.238$; five-fold cross-validated $R^2=.216\pm.067$; Figure~\ref{fig:coefs}).
Coefficient magnitudes indicate that submission count and cross-competition engagement dominate the model, while discussion-based metrics contribute little once productive engagement is included.
This pattern reinforces the distinction between visible social participation and hands-on competitive practice.
Users may post in megathreads for many reasons, including networking, reassurance, or curiosity, but the leaderboard ultimately rewards repeated experimentation with models and data.
The results are therefore consistent with an interpretation in which productive engagement is closer to the platform's core learning mechanism than megathread conversation is.

\subsection{RQ4: Expertise as Moderator}

\begin{figure}[t]
\Description{Two box plots by tier, showing competition performance (100 minus percentile) in the left panel and submission count in the right panel. Median performance is about 57 for Novices, 46 for Contributors, 69 for Experts, and 85 for Masters. Each performance group is annotated with its sample size and mean. The submission-count panel shows the distributions of submission activity across tiers.}
\centering
\includegraphics[width=\textwidth]{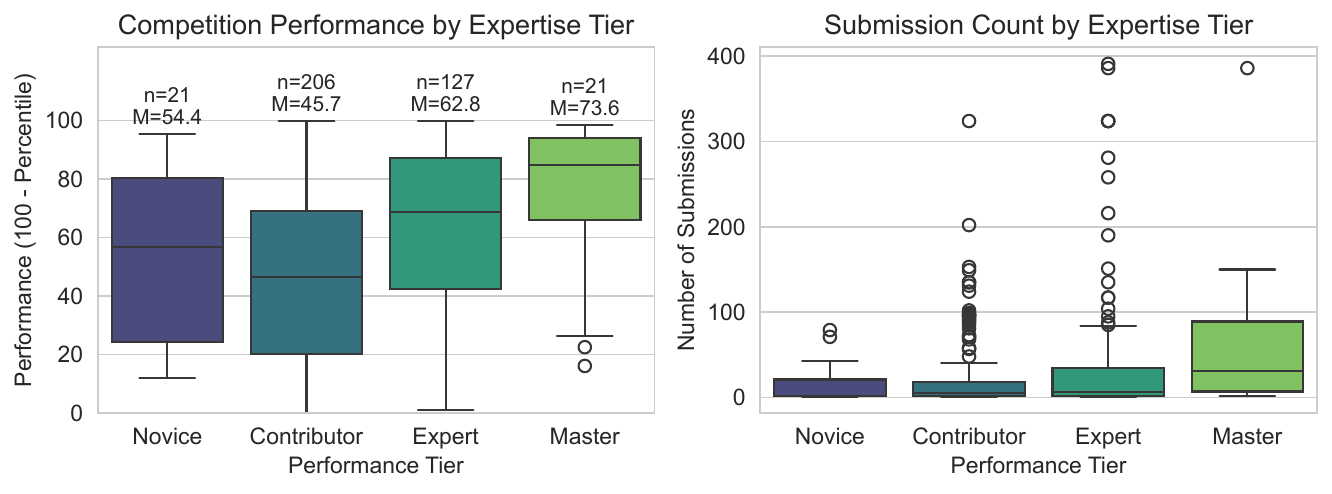}
\caption{Competition performance (left) and submission count (right) by tier (Study~2). Higher-tier users achieve substantially better outcomes (KW $H=37.80$, $p<.001$).}
\label{fig:perf_tier}
\end{figure}

Finally, we ask whether expertise changes \emph{how} participation converts into performance, not just whether higher tiers score better overall.
Performance differs significantly across tiers (Kruskal-Wallis $H=37.80$, $p<.001$; Figure~\ref{fig:perf_tier}).
Master-tier users achieve the highest median performance (84.8) and Experts the next highest (68.7), both above the Contributor median (46.5); the Novice subsample is small ($n=21$), with a median of 56.7.
Pairwise Mann-Whitney $U$ tests confirm that Experts and Masters each significantly outperform Contributors (both $p<.001$), whereas the Expert--Novice difference is not significant given the small Novice subsample ($p=.20$).

Submission count and performance are positively associated within every tier, but precision differs substantially across subgroups (Table~\ref{tab:mod}).
The Novice estimate is positive but not statistically significant ($\rho=.322$, $p=.154$, $n=21$); a post-hoc power analysis indicates that this subgroup's power to detect a medium-sized effect is only $1-\beta\approx.29$, well below the .80 convention.
The Contributor, Expert, and Master estimates are all significant.
The monotonic pattern is therefore suggestive rather than conclusive evidence that the submission-performance association strengthens with expertise.
A Fisher $z$-transformation comparing $\rho_{\text{Novice}}=.322$ and $\rho_{\text{Master}}=.675$ yields $z=1.46$, $p=.145$, confirming that the difference between the extreme subgroups does not reach significance with the present sample sizes.
This interpretation is broadly consistent with a Communities of Practice account~\cite{lave:lpp}, in which members closer to the center of the practice may extract more value from each feedback cycle.
At the same time, hierarchical regression shows that tier adds only marginal $R^2$ (+.003, $p=.216$), tempering any simple circular interpretation and indicating that visible expertise does not independently explain much additional variance beyond participation behaviors.
Substantively, this complicates a simple ``experts win because they are experts'' story.
What appears to vary across tiers is not only absolute performance, but also how efficiently practice may be converted into gains.
Experienced users may recognize stronger ideas sooner, diagnose leaderboard movement more effectively, or avoid low-value experimentation.
Consistent with this reading, the submission-performance correlation strengthens across tiers (from a non-significant $\rho=.322$ among Novices to $\rho=.675$ among Masters). Performance rises more consistently with submission activity among more experienced users, suggesting that expertise operates less as a fixed advantage than as a multiplier on the value extracted from each feedback cycle.
At the same time, the small Novice and Master subsamples mean that the present data are better suited to identifying a plausible pattern than to establishing a precise tier-by-tier functional form.

\begin{table}[t]
\centering
\small
\caption{Tier-stratified Spearman correlations between submission count and performance (Study~2). The increasing pattern is suggestive, but the Novice estimate is not significant.}
\begin{tabular*}{\linewidth}{@{\extracolsep{\fill}}lcccc@{}} \toprule
\textbf{Statistic} & \textbf{Novice} & \textbf{Contributor} & \textbf{Expert} & \textbf{Master} \\ \midrule
$n$                 & 21      & 206     & 127     & 21      \\
$\boldsymbol{\rho}$ & $+$.322 & $+$.351 & $+$.476 & $+$.675 \\
$\boldsymbol{p}$    & .154    & $<$.001 & $<$.001 & $<$.001 \\
\bottomrule
\end{tabular*}
\label{tab:mod}
\end{table}

\section{Discussion}
In this paper, we investigate how expertise shapes team formation and performance in Kaggle's L4T megathreads. We found that experts received higher social attention and stronger competition performance than the large contributor majority, while novices were more numerous but received less attention. Relatedly, we found that productive engagement (submissions) was more predictive of performance than discussion-based participation. These findings have implications for understanding how online communities support learning and collaboration, and how platform design can influence who benefits from participation.
The discussion unfolds in that order. We first interpret the expert--novice attention gap, then the divide between social coordination and productive practice, before turning to what these results imply for how collaboration is measured, what they suggest for design and research, and the limitations that bound our claims.

\subsection{The Expert-Novice Paradox: Access without Equal Participation}
Megathreads are overwhelmingly used by low-tier users (novices and contributors; 89.3\%) yet the most visible benefits accrue disproportionately to experienced users.
Study~1 shows that higher-tier users receive more incoming replies and higher forum-visible teaming success rates despite being a numerical minority in the community.
This pattern matters because the megathread offers access to collaboration. It lowers the cost of finding collaborators by placing everyone in the same public thread.
In practice, however, the thread does not erase existing differences.
Instead, visible expertise appears to shape who is noticed, who is trusted, and who is more likely to convert attention into reciprocal interaction.

This is the paradox of the platform. It attracts newcomers, but newcomer-heavy participation does not translate into newcomer advantage.
From a Communities of Practice perspective, L4T megathreads function as access to the community, yet such access does not promise equal participation.
Legitimate peripheral participation assumes that novices can move toward fuller participation through access and interaction~\cite{lave:lpp}. However, our results suggest that access remains socially filtered by visible expertise.
Experts bring not only skill but also legible proof of skill~\cite{cheng2020building}.
That proof appears to matter in a setting where strangers must make rapid judgments about potential teammates.

The trust-signal analysis suggests how this filtering operates in practice.
When collaboration begins among strangers, participants judge credibility from whatever cues are at hand~\cite{marlow:impression}.
Higher-tier users can point to platform-verified achievement, a signal that is costly to fake and immediately legible on the platform~\cite{ren2020show}, although explicit references to past performance are uncommon in every tier.
In warranting terms, tier badges and medal histories are system-generated cues whose credibility observers have little reason to discount~\cite{walther:warranting}.
Novices, lacking such records, compensate with self-generated, lower-warrant signals, emphasizing motivation, credentials, and offers to continue contact off-platform.
This asymmetry in self-presentation is consistent with evidence from other team-formation settings that novices and experienced members weigh different criteria when selecting collaborators~\cite{das2023comparative}.
It also sharpens the paradox. The attention gap is not a matter of effort or silence, because novices are actively signaling, but the signals available to them are harder to verify and appear to carry less weight in a market of strangers.

At the same time, Kaggle tier is based on cumulative historical success, so any claim that experts do better risks sounding circular.
Our results support a narrower conclusion.
Tier is a cross-competition achievement signal, whereas Study~2 measures performance within a particular competition on held-out test data.
More importantly, hierarchical regression shows that tier adds only marginal explanatory power beyond participation variables ($\Delta R^2=.003$, $p=.216$).
Our contribution is therefore not the tautology that historically successful users succeed again.
It is that visible expertise is associated with unequal social attention in team-seeking spaces, and that participation appears to pay off differently across expertise levels.

These findings carry a message back to both literatures we build on.
For Communities of Practice theory, they suggest a boundary condition on legitimate peripheral participation.
LPP treats access as the gateway through which newcomers begin moving toward fuller participation~\cite{lave:lpp}, but our results indicate that in platform-mediated communities, where reputation infrastructure makes each member's history publicly legible, access itself becomes graded. The periphery is not a shared starting line but a sorted queue in which visible status shapes who is engaged further.
CoP researchers studying online communities should therefore treat access as a stratified resource conditioned by the platform's reputation design, rather than as a binary property of open membership.
For group formation research, the findings extend the self-presentation and information-availability themes~\cite{harris:groupformation} to a naturalistic competitive setting. Among strangers who lack the prior social ties that users otherwise fall back on~\cite{gomez2019would}, system-generated status cues appear to organize attention instead, concentrating it on already-visible members.
The broader lesson is to examine not only whether a platform feature works, but also for whom it works best.
Aggregate participation counts can make a community look inclusive even when opportunity is unevenly distributed once users are disaggregated by expertise.
In other words, open access can coexist with stratified visibility.
That tension is likely relevant beyond Kaggle, especially in online learning settings where reputational traces are highly public~\cite{wang2020marginality}.

\subsection{The Gap Between Coordination and Practice}
Navigating this social sorting and forming a group is only the first step. Once users enter the competition space, the nature of their participation must shift to drive competitive or collaborative success. Our analysis reveals a distinct boundary between social coordination and productive participation. Megathread-specific discussion does not predict competition performance, whereas productive engagement, submissions, does.
This finding is important because discussion activity is often used as a convenient proxy for engagement in online platforms~\cite{zhang2018characterizing}.
Our results suggest that such a proxy is too coarse in competitive environments like Kaggle.

The moderation analysis (RQ4) adds nuance to this picture.
The submission-performance correlation strengthens monotonically across tiers, from $\rho=.322$ among Novices to $\rho=.675$ among Masters.
From a Communities of Practice perspective, this is what a trajectory from peripheral to full participation would predict~\cite{lave:lpp}.
Members closer to the center of the practice have internalized the community's shared repertoire, such as how to read leaderboard movement, which ideas are worth pursuing, and what a promising experiment looks like.
Submitting a model is therefore not the same activity at every position in the community; the same observable action draws on different levels of accumulated, largely tacit understanding.
Newcomers, by contrast, may perform the visible behaviors of participation without yet having access to the interpretive resources that make each feedback cycle informative.
This reading also connects the two studies. Novices appear peripheral both socially, receiving less attention in the megathread, and epistemically, extracting less from each submission.
However, as reported above, the Novice estimate is not statistically significant, that subgroup is underpowered, and the difference between the extreme tiers does not reach significance.
We therefore treat the monotonic pattern as suggestive rather than conclusive. It identifies a plausible mechanism, participation that converts into competence at different rates across positions in the community, which future work with larger novice samples should test directly.
If it holds, the pattern refines the Communities of Practice account of participation-as-learning. Participation is not a single currency, and its value for competence development depends jointly on its form, discussion versus practice, and on the participant's position in the community.

By contrast, the primary function of L4T megathreads is social coordination, not technical knowledge building.
Many posts are short requests for teammates, contact details, or brief self-introductions.
Those interactions may still matter for belonging, opportunity, and future collaboration, but they are not designed to deliver the same feedback-rich learning cycle as submitting solutions.
This helps explain why megathread posting alone is not strongly linked to private leaderboard outcomes.
The conclusion is not that discussion never matters, but that team-seeking discussion, by itself, is a weak proxy for competitive success in this setting.

This distinction is especially relevant for social computing research that relies on behavioral traces, where readily available data can tempt us to equate communication volume with meaningful participation.
In environments where performance depends on iterative production, behavioral traces tied to actual practice may be more informative than discussion traces alone~\cite{wise:participation, siemens:la}.

\subsection{Reevaluating How We Measure Collaboration}
As online collaboration environments increasingly blur the lines between discussion forums and workspaces, it is no longer enough to ask whether a community feature increases participation overall. Our combination of reply networks, verified team rosters, and private leaderboard rankings illustrates the necessity of separating the social act of team formation from the productive act of competing.

Leaderboard rankings offer unusually concrete, standardized proxies for performance in an open collaboration environment~\cite{siemens:la}, moving the analysis beyond self-reported outcomes. More importantly, this approach reveals that the most visible traces are not necessarily the most collaboration-relevant ones. Forum posts are easy to count, but in this setting, they tell a very different story than model submissions do.
This is methodologically valuable because many studies of online participation rely on self-report, course completion, or broad participation counts.
Kaggle provides a different opportunity, because process traces, public interaction data, and objective outcomes coexist on the same platform.

Our approach combines reply networks, competition performance, and verified team rosters, three sources that are rarely analyzed together.
The roster linkage is especially important because it moves the analysis beyond pure discourse traces and allows partial validation of whether public team-seeking behavior corresponds to realized collaboration.
Even though the validation is incomplete, the combination of network, content, and outcome data makes it possible to study not only who interacts, but also how visible interaction relates to later teaming and performance.
It illustrates the value of connecting social traces to consequential outcomes rather than analyzing each trace stream in isolation.
The broader methodological lesson is that convenience and validity are not the same thing. The best behavioral proxy depends on the platform's actual participation and coordination mechanism.

\subsection{Implications for Design and Research}
Our findings suggest that L4T megathreads should be understood not as neutral discussion spaces but as sociotechnical filters that shape the visibility of potential collaborators.
Users are not choosing teammates from nowhere.
They are choosing through interfaces that foreground some signals, hide others, and reward certain histories of participation~\cite{harris:groupformation, marlow:impression}.
This framing turns the disparities documented in Study~1 into a design problem. If visibility is structured by the interface, then the interface is also a lever.

The most direct lever concerns the importance of reputation cues.
Because tier is highly visible, users may over-rely on it when deciding whom to contact~\cite{gomez2019would}.
Completely hiding expertise would reduce useful information and make coordination harder, while leaving status cues fully unmediated may reproduce existing hierarchies~\cite{wang2020marginality}.
The design challenge is therefore not whether to show expertise, but how to prevent it from becoming the only cue that matters.
Platforms could experiment with delaying or softening tier visibility in early matching stages, or with foregrounding complementary skills, recent activity, and responsiveness alongside tier badges so that users are evaluated on a broader profile than reputation alone.
Such nudges would align with prior work on supporting self-formed crowd teams~\cite{vinella:nudging}.

A second lever is structured self-presentation.
Our trust-signal analysis shows that novices already signal actively, but through low-cost cues that are hard to verify.
Templates that elicit the information currently scattered across ad hoc posts, such as preferred tasks, coding strengths, time availability, and prior relevant experience, would not remove status differences, but they could reduce the burden on newcomers to guess which details signal credibility~\cite{ren2020show}.

Third, platforms should support \emph{conversion}, not just contact.
Current megathreads are good at making requests visible, but much less structured when it comes to helping users evaluate fit, establish working norms, or transition into actual collaboration.
Because the strongest performance signals come from practice rather than discussion, team-seeking tools should make it easier for newly formed groups to move from conversation to joint work, for example through shared notebook spaces, prompts to state expected contribution levels, and short checklists that help potential teammates compare goals and commitment before moving off-thread~\cite{lykourentzou2017team, salehi2017huddler}.
Novice-focused diagnostic supports could further help users interpret leaderboard feedback and convert participation into learning~\cite{kizilcec:equity}.
Such scaffolding would be especially useful for novices, who may lack both status signals and prior experience in negotiating collaborative expectations online.
The evidence does not prove that these interventions will succeed, but it indicates where inequities currently arise and therefore where design attention is most warranted.

These implications extend beyond Kaggle.
Many online collaboration environments are now hybrid spaces in which social interaction, work production, and public reputation are tightly interwoven, from open-source communities to online creator ecosystems and project-based learning platforms~\cite{marlow:impression, cheng2020building}.
In such settings, researchers and designers must ask which users become more legible, which forms of contribution are rewarded, and whether platform mechanisms widen or narrow access to meaningful collaboration.

\subsection{Limitations and Future Work}
Three limitations matter most for interpreting our claims.
First, our measures are imperfect proxies for the constructs they represent.
Kaggle tier captures cumulative platform achievement but also reflects tenure and prior opportunity, so we interpret it as a visible platform signal rather than a pure measure of underlying ability.
The trust-signal analysis is keyword-based and descriptive, less precise than full manual coding.
And our participation measures omit behaviors such as notebook forking, lurking, and off-platform collaboration, which may matter most for users who improve by observing rather than posting; likewise, leaderboard placement is not the only outcome participants value.

Second, the Study~1 success outcome captures forum-visible teaming traction rather than verified team formation.
The roster linkage partially supports the proxy, showing that 26 verified teams included at least two megathread posters, but it does not establish that every reciprocal reply led to a team, and some collaboration likely moved to email, Discord, or other private channels.
Longitudinal tracking of how public interaction turns into realized, and potentially durable, teams is the most direct way to close this gap.

Finally, the study is observational and cross-sectional, and Study~2 contains repeated user-competition observations, so our results should be read as descriptive associations. We cannot infer that megathread participation causes performance gains, or that visible expertise causes others to reply.
Future work could strengthen the statistics with multilevel models or cluster-robust inference, probe mechanisms through interviews with novice and expert participants, and test causal questions directly through natural experiments in how tier information is displayed or quasi-experimental designs around promotion thresholds.
Such work would move the literature from diagnosis toward intervention.

\subsection{Ethical Considerations}

We use publicly available Kaggle API data; all identifiers are anonymized and we report only tier-level aggregates.
By operationalizing success through leaderboard rankings, we adopt the platform's framing, and we recognize that public trace data can make participants appear more fully legible than they really are. Tier, comments, and team rosters reveal only the platform-visible portion of collaboration, and users may coordinate privately, work under unequal time constraints, or derive value (social connection, belonging, occupational experience) that never appears in leaderboard metrics.
For this reason, we avoid reporting identifiable case narratives and interpret public traces as partial indicators rather than complete portraits of participant intent or capability.

\section{Conclusion}

Combining network analysis, roster validation, and participation-performance modeling, we examined how expertise shapes collaboration and achievement in Kaggle's 2023 L4T megathreads.
Study~1 shows that the forum is novice-heavy but attention is unevenly distributed. Higher-tier users receive more replies and higher rates of forum-visible teaming success.
The roster linkage indicates that megathread participation sometimes corresponds to realized teams, while also showing that public interaction should not be equated too quickly with verified team formation.
Study~2 shows that productive engagement is more strongly associated with leaderboard performance than megathread discussion alone, and that this relationship appears stronger among higher-tier users, albeit with imprecision in the smallest subgroups.

The findings suggest that platform features designed to open access do not automatically flatten hierarchy.
Visible expertise can structure who gets attention, who appears trustworthy, and who seems best positioned to convert participation into competitive success.
The paper highlights two broader lessons for team formation study. First, equity questions should be asked at the subgroup level rather than assumed from aggregate participation. Second, behavioral proxies should be matched to the underlying practice of the platform.
In online communities that host data science competitions like Kaggle, what participants do may be more diagnostic than what they say.

\begin{acks}
During the preparation of this work, the authors used ChatGPT for language editing purposes to improve readability, and OpenAI's GPT-5.3-codex to assist in converting the manuscript from a non-ACM format into the required ACM template.
The tools were not used to generate any core ideas, concepts, or references.
After using these services, all authors carefully reviewed and edited the manuscript under strict human supervision, and take full responsibility for the original ideas and final content of the published article.

\end{acks}

\bibliographystyle{ACM-Reference-Format}
\bibliography{references}

@book{wenger:cop,
  title={Communities of Practice: Learning, Meaning, and Identity},
  author={Wenger, Etienne},
  year={1998},
  publisher={Cambridge University Press},
  doi={10.1017/cbo9780511803932}
}

@book{lave:lpp,
  title={Situated Learning: Legitimate Peripheral Participation},
  author={Lave, Jean and Wenger, Etienne},
  year={1991},
  publisher={Cambridge University Press},
  doi={10.1017/cbo9780511815355}
}

@incollection{marsick:informal,
  title={Informal and incidental learning in the workplace},
  author={Marsick, Victoria J. and Watkins, Karen E. and Callahan, Mary Wilson and Volpe, Marie},
  booktitle={Handbook of Research on Adult Learning and Development},
  editor={Smith, M. Cecil and DeFrates-Densch, Nancy},
  pages={570--599},
  year={2009},
  publisher={Routledge}
}

@article{barabasi:preferential,
  title={Emergence of scaling in random networks},
  author={Barab\'{a}si, Albert-L\'{a}szl\'{o} and Albert, R\'{e}ka},
  journal={Science},
  volume={286},
  number={5439},
  pages={509--512},
  year={1999},
  doi={10.1126/science.286.5439.509}
}

@article{mcpherson:homophily,
  title={Birds of a feather: Homophily in social networks},
  author={McPherson, Miller and Smith-Lovin, Lynn and Cook, James M.},
  journal={Annual Review of Sociology},
  volume={27},
  number={1},
  pages={415--444},
  year={2001},
  doi={10.1146/annurev.soc.27.1.415}
}

@article{huang:crowdsourcing,
  title={Being a solo endeavor or team worker in crowdsourcing contests? It is a long-term decision you need to make},
  author={Huang, Keman and Zhou, Jilei and Chen, Shao},
  journal={Proceedings of the ACM on Human-Computer Interaction},
  volume={6},
  number={CSCW2},
  articleno={494},
  numpages={32},
  year={2022},
  doi={10.1145/3555595}
}

@article{twyman:kaggle,
  title={Positioning in a collaboration network and performance in competitions: A case study of {Kaggle}},
  author={Twyman, Marlon and Muri\'{c}, Goran and Zheng, Weiwei},
  journal={Journal of Computer-Mediated Communication},
  volume={28},
  number={4},
  pages={zmad024},
  year={2023},
  doi={10.1093/jcmc/zmad024}
}

@inproceedings{dissanayake:crowdsourcing,
  title={Team formation and performance in online crowdsourcing competitions: The role of homophily and diversity in solver characteristics},
  author={Dissanayake, Indika and Nerur, Sridhar and Zhang, Jie},
  booktitle={Proceedings of the 40th International Conference on Information Systems (ICIS 2019)},
  publisher={Association for Information Systems},
  address={Munich, Germany},
  year={2019}
}

@article{li:kaggle,
  title={Exploring the social influence of the {Kaggle} virtual community on the {M5} competition},
  author={Li, Xixi and Bai, Yun and Kang, Yanfei},
  journal={International Journal of Forecasting},
  volume={38},
  number={4},
  pages={1507--1518},
  year={2022},
  doi={10.1016/j.ijforecast.2021.10.001}
}

@inproceedings{chow:kaggle,
  title={A pedagogy that uses a {Kaggle} competition for teaching machine learning: An experience sharing},
  author={Chow, Winn},
  booktitle={2019 IEEE International Conference on Engineering, Technology and Education (TALE)},
  pages={1--5},
  year={2019},
  publisher={IEEE},
  doi={10.1109/tale48000.2019.9226005}
}

@article{polak:kaggle,
  title={A study on student performance, engagement, and experience with {Kaggle} {InClass} data challenges},
  author={Polak, Julia and Cook, Dianne},
  journal={Journal of Statistics and Data Science Education},
  volume={29},
  number={1},
  pages={63--70},
  year={2021},
  doi={10.1080/10691898.2021.1892554}
}

@article{wise:participation,
  title={Attending to others' posts in asynchronous discussions: Learners' online ``listening'' and its relationship to speaking},
  author={Wise, Alyssa Friend and Hausknecht, Simone Nicole and Zhao, Yuting},
  journal={International Journal of Computer-Supported Collaborative Learning},
  volume={9},
  number={2},
  pages={185--209},
  year={2014},
  doi={10.1007/s11412-014-9192-9}
}

@article{harris:groupformation,
  title={Joining together online: The trajectory of {CSCW} scholarship on group formation},
  author={Harris, Alexa M. and G\'{o}mez-Zar\'{a}, Diego and DeChurch, Leslie A. and Contractor, Noshir S.},
  journal={Proceedings of the ACM on Human-Computer Interaction},
  volume={3},
  number={CSCW},
  articleno={148},
  numpages={27},
  year={2019},
  doi={10.1145/3359250}
}

@inproceedings{marlow:impression,
  title={Impression formation in online peer production: Activity traces and personal profiles in {GitHub}},
  author={Marlow, Jennifer and Dabbish, Laura and Herbsleb, Jim},
  booktitle={Proceedings of the 2013 Conference on Computer Supported Cooperative Work},
  pages={117--128},
  year={2013},
  publisher={Association for Computing Machinery},
  address={New York, NY, USA},
  doi={10.1145/2441776.2441792}
}

@incollection{walther:warranting,
  title={Cues filtered out, cues filtered in: Computer-mediated communication and relationships},
  author={Walther, Joseph B. and Parks, Malcolm R.},
  booktitle={Handbook of Interpersonal Communication},
  editor={Knapp, Mark L. and Daly, John A.},
  edition={3rd},
  pages={529--563},
  year={2002},
  publisher={Sage Publications},
  address={Thousand Oaks, CA}
}

@inproceedings{vinella:nudging,
  title={The impact of digital nudging techniques on the formation of self-assembled crowd project teams},
  author={Vinella, Federica Lucia and Mosch, Rosa and Lykourentzou, Ioanna and Masthoff, Judith},
  booktitle={Proceedings of the 30th ACM Conference on User Modeling, Adaptation and Personalization},
  pages={265--275},
  year={2022},
  publisher={Association for Computing Machinery},
  address={New York, NY, USA},
  doi={10.1145/3503252.3531298}
}

@article{siemens:la,
  title={Learning analytics: The emergence of a discipline},
  author={Siemens, George},
  journal={American Behavioral Scientist},
  volume={57},
  number={10},
  pages={1380--1400},
  year={2013},
  doi={10.1177/0002764213498851}
}

@article{kizilcec:equity,
  title={Scaling up behavioral science interventions in online education},
  author={Kizilcec, Ren\'{e} F. and Reich, Justin and Yeomans, Michael and Dann, Christoph and Brunskill, Emma and Lopez, Glenn and Turkay, Selen and Williams, Joseph J. and Tingley, Dustin},
  journal={Proceedings of the National Academy of Sciences of the United States of America},
  volume={117},
  number={26},
  pages={14900--14905},
  year={2020},
  doi={10.1073/pnas.1921417117}
}

@inproceedings{wei:megathread,
  title={Social network and self-representation in megathread: Group formation in a data science crowdsourcing community},
  author={Wei, Shiyao and Bi, Ran},
  booktitle={Proceedings of the 17th International Conference on Educational Data Mining},
  pages={738--743},
  year={2024},
  publisher={International Educational Data Mining Society},
  doi={10.5281/zenodo.12729934}
}

@inproceedings{bi2026benefits,
  title={Who benefits from looking-for-team forums? Examining expert-novice differences in online team formation},
  author={Bi, Ran and Wei, Shiyao},
  booktitle={Proceedings of the 19th International Conference on Educational Data Mining},
  pages={813--817},
  year={2026},
  publisher={International Educational Data Mining Society},
  doi={10.5281/zenodo.21040055}
}

@inproceedings{hagberg:networkx,
  title={Exploring network structure, dynamics, and function using {NetworkX}},
  author={Hagberg, Aric A. and Schult, Daniel A. and Swart, Pieter J.},
  booktitle={Proceedings of the 7th Python in Science Conference},
  pages={11--15},
  year={2008},
  publisher={SciPy},
  doi={10.25080/tcwv9851}
}

@article{scikit-learn,
  title={Scikit-learn: Machine learning in {Python}},
  author={Pedregosa, Fabian and Varoquaux, Ga\"{e}l and Gramfort, Alexandre and Michel, Vincent and Thirion, Bertrand and Grisel, Olivier and Blondel, Mathieu and Prettenhofer, Peter and Weiss, Ron and Dubourg, Vincent and Vanderplas, Jake and Passos, Alexandre and Cournapeau, David and Brucher, Matthieu and Perrot, Matthieu and Duchesnay, \'{E}douard},
  journal={Journal of Machine Learning Research},
  volume={12},
  number={85},
  pages={2825--2830},
  year={2011}
}

@article{cheng2020building,
  title={Building community knowledge in online competitions: Motivation, practices and challenges},
  author={Cheng, Ruijia and Zachry, Mark},
  journal={Proceedings of the ACM on Human-Computer Interaction},
  volume={4},
  number={CSCW2},
  articleno={179},
  numpages={22},
  year={2020},
  publisher={Association for Computing Machinery},
  address={New York, NY, USA},
  doi={10.1145/3415250}
}

@inproceedings{gomez2019would,
  title={Who would you like to work with? {Use} of individual characteristics and social networks in team formation systems},
  author={G{\'o}mez-Zar{\'a}, Diego and Paras, Matthew and Twyman, Marlon and Lane, Jacqueline N. and DeChurch, Leslie A. and Contractor, Noshir S.},
  booktitle={Proceedings of the 2019 CHI Conference on Human Factors in Computing Systems},
  pages={1--15},
  year={2019},
  publisher={Association for Computing Machinery},
  address={New York, NY, USA},
  doi={10.1145/3290605.3300889}
}

@inproceedings{lykourentzou2017team,
  title={Team dating leads to better online ad hoc collaborations},
  author={Lykourentzou, Ioanna and Kraut, Robert E. and Dow, Steven P.},
  booktitle={Proceedings of the 2017 ACM Conference on Computer Supported Cooperative Work and Social Computing},
  pages={2330--2343},
  year={2017},
  publisher={Association for Computing Machinery},
  address={New York, NY, USA},
  doi={10.1145/2998181.2998322}
}

@inproceedings{salehi2017huddler,
  title={Huddler: Convening stable and familiar crowd teams despite unpredictable availability},
  author={Salehi, Niloufar and McCabe, Andrew and Valentine, Melissa and Bernstein, Michael},
  booktitle={Proceedings of the 2017 ACM Conference on Computer Supported Cooperative Work and Social Computing},
  pages={1700--1713},
  year={2017},
  publisher={Association for Computing Machinery},
  address={New York, NY, USA},
  doi={10.1145/2998181.2998300}
}

@article{zhang2018characterizing,
  title={Characterizing online public discussions through patterns of participant interactions},
  author={Zhang, Justine and Danescu-Niculescu-Mizil, Cristian and Sauper, Christina and Taylor, Sean J.},
  journal={Proceedings of the ACM on Human-Computer Interaction},
  volume={2},
  number={CSCW},
  year={2018},
  publisher={Association for Computing Machinery},
  address={New York, NY, USA},
  articleno={198},
  numpages={27},
  doi={10.1145/3274467}
}

@article{javadi2024contribution,
  title={Contribution to team and community in crowdsourcing contests: A qualitative investigation},
  author={Javadi Khasraghi, Hanieh and Vaghefi, Isaac and Hirschheim, Rudy},
  journal={Information Technology \& People},
  volume={37},
  number={1},
  pages={223--250},
  year={2024},
  publisher={Emerald Publishing Limited},
  doi={10.1108/itp-01-2021-0069}
}

@article{blohm2011does,
  title={Does collaboration among participants lead to better ideas in IT-based idea competitions? An empirical investigation},
  author={Blohm, Ivo and Bretschneider, Ulrich and Leimeister, Jan Marco and Krcmar, Helmut},
  journal={International Journal of Networking and Virtual Organisations},
  volume={9},
  number={2},
  pages={106--122},
  year={2011},
  publisher={Inderscience Publishers},
  doi={10.1504/ijnvo.2011.042413}
}

@article{ren2020show,
  title={Show me your expertise before teaming up: Sharing online profiles predicts success in open innovation},
  author={Ren, Ruqin and Yan, Bei and Jian, Lian},
  journal={Internet Research},
  volume={30},
  number={3},
  pages={845--868},
  year={2020},
  publisher={Emerald Publishing Limited},
  doi={10.1108/intr-01-2019-0024}
}

@article{das2023comparative,
  title={A comparative study of novice and habitual entrepreneur's choice for founding team member},
  author={Das, Willy and Das, Satyasiba},
  journal={Journal of Entrepreneurship in Emerging Economies},
  volume={15},
  number={6},
  pages={1373--1391},
  year={2023},
  publisher={Emerald Publishing Limited},
  doi={10.1108/jeee-12-2021-0456}
}

@article{wang2020marginality,
  title={Marginality and team building in collaborative crowdsourcing},
  author={Wang, Rong},
  journal={Online Information Review},
  volume={44},
  number={4},
  pages={827--846},
  year={2020},
  publisher={Emerald Publishing Limited},
  doi={10.1108/oir-09-2018-0269}
}

\end{document}